\documentclass[twocolumn]{article}
\usepackage[utf8]{inputenc}

\RequirePackage[breakable, most]{tcolorbox}
\RequirePackage{float}
\RequirePackage{xcolor}

\definecolor{LightBlue}{HTML}{3389f8}

\usepackage[margin=1in]{geometry}

\usepackage{graphicx}   

\usepackage{url}        

\usepackage{amsmath}    

\usepackage{caption}
\usepackage{subcaption} 
\usepackage{multicol}

\usepackage{float} 

\definecolor{main}{HTML}{5989cf}    
\definecolor{sub}{HTML}{cde4ff}     

\newtcolorbox{box_blue}{
    enhanced,
    boxrule = 0pt,
    colback = sub,
    borderline west = {1pt}{0pt}{main}, 
    borderline west = {0.75pt}{2pt}{main}, 
    borderline east = {1pt}{0pt}{main}, 
    borderline east = {0.75pt}{2pt}{main},
    borderline north = {1pt}{0pt}{main},
    borderline north = {0.75pt}{2pt}{main},
    borderline south = {1pt}{0pt}{main},
    borderline south = {0.75pt}{2pt}{main},
}

\usepackage{ifthen}

\newboolean{showpartopic}
\setboolean{showpartopic}{true} 

\newcommand{\partopic}[1]{%
  \ifthenelse{\boolean{showpartopic}}{\textbf{#1}}{}%
}

\usepackage[backend=biber,sorting=none,style=phys]{biblatex} 

\usepackage{authblk}

\usepackage{abstract}

\onecolumn

\title{Hackathons for biophysics education: simulating the cytoskeleton}

\date{\vspace{-5ex}} 

\author[1,2]{Yoav G. Pollack\thanks{yoavpol@gmail.com}}
\author[1]{Komal Bhattacharyya}
\author[1]{Anas Hussin}
\author[1]{Emily Klass}
\author[1]{Raffaele Mendozza}
\author[1,4]{Krishna Iyer V S}
\author[1,3]{Gerrit Wellecke}
\author[1,3]{Patrick Zimmer}
\affil[1]{University of Göttingen, Göttingen, Germany.}
\affil[2]{Max Planck Institute for Multidisciplinary Sciences, Göttingen, Germany.}
\affil[3]{Max Planck Institute for Dynamics and Self--Organization, Göttingen, Germany.}
\affil[4]{Indian Institute of Science Education and Research, Pune, India.}
\begin{document}
\twocolumn[
    \maketitle
    \begin{onecolabstract}
        Hackathons are intensive innovation-oriented events where participants work in teams to solve problems or create projects in as little as 24 or 48 hours. These events are common in startup culture, open source communities and mainstream industry. Here we examine how hackathons can be ported to academic teaching, specifically in computational biophysics. We propose hackathons as a teaching modality distinct from traditional courses and structured workshops. In particular, we suggest they can offer a low-stakes platform for students to overcome entry barriers to computational tools or to explore new topics, disciplines, and skills beyond their academic comfort zone. We tested this format in two computational biophysics hackathons on the Göttingen campus in 2023 and 2024, providing practical insights and a preliminary evaluation. To the best of our knowledge, the 2024 event is the first \textit{public} hackathon dedicated to Biophysics. This paper explores the benefits of the hackathon format for teachers and researchers and provides guidelines for running a hackathon adapted to a teaching goal.
        \vspace{0.5cm}
    \end{onecolabstract}
]
\saythanks
\section{Introduction}

 Traditional academic courses cover content in depth, while intensive courses and workshops can effectively teach practical know-how. Such structured platforms are well suited for surpassing a given competency threshold, for example as part of academic qualification programs or for training in specific scientific tools. However, a low-stakes ``trying out" of a new topic or computational tool is often not compatible with these more rigid teaching modalities; an unfinished course is typically perceived as ``wasted time" rather than a minimal learning package. In line with the steadily growing trend of project-oriented and extracurricular learning modalities\cite{willis2017challenge} we propose the hackathon format as a method aimed at getting learners just over initial entry barriers to a point enabling independent exploration.

\partopic{What is a hackathon?}
Hackathons are time-constrained innovation-driven events where technology enthusiasts collaborate or compete to develop computational projects\cite{briscoe2014digital}. These events typically last 24 to 48 hours, often continuously. Participants form small groups to create functional prototypes addressing a pre-defined problem and showcase their projects to experts and the audience. Other hackathons are more open-ended with a common topical or tool-oriented theme (e.g. Telecom industry or Machine Learning).
The hackathon concept is widely used to introduce innovation into (often big and rigid) organizations such as corporations\cite{pe-than_designing_2019} or public sector institutions \cite{yuan2021open}. Companies sometimes use hackathons for more quantifiable goals such as talent recruitment, or developing minimal products as a replacement for official outsourcing\cite{nolte_you_2018, kalleberg_hackathons_2017}. Hackathons are also prevalent in open-source and academic communities\cite{ chandrasekaran_best_2018, maaravi_running_2018, gama_hackathon_2018, kienzler_learning_2017, nandi2016hackathons, falk2024future}, mostly in computer science contexts.
While learning is reported in surveys as the number one reason for attending a hackathon, the vast majority of hackathon goers also report themselves as already possessing relevant technical skills\cite{briscoe2014digital} suggesting the typical use case is to further existing knowhow.

\partopic{The teaching-hackathon.} Here we argue that the compacted time format together with the inherent lack of a pre-known solution, create an ideal ``low-stakes" scenario attracting students to try out new computational tools; the perceived lack of a strict success measure actually constituting an advantage. In addition, the inherently unorthodox, and ideally fun, event atmosphere can motivate students to explore new scientific topics and disciplines. Finally, the constrained event duration can mean an affordable time commitment between more lengthy high-priority study and research undertakings.

 We however suggest some adaptations to the hackathon format when the declared goal is getting learners over entry barriers (as opposed to end-product driven hackathons). Traditional hackathons are geared towards development by participants with a strong technical foundation and learning is left to happen organically and spontaniously\cite{schulten2024we}. Here however, we propose a minimal amount of guided learning; just enough to cross the entry barrier, while still reserving most of the time to the ``learn by trying" style. Finding this balance is not trivial\cite{schulten2024we}, and we discuss this below. This new teaching hackathon methodology was tested in two (retrospectively successful) trials as detailed below, which provided useful practical insights and preliminary evaluation of the format's potential.

\section{Methods}
\label{sec:methods}

\partopic{Trial hackathons.} 
Two biophysics oriented hackathons were organized in the Göttingen (Gö) campus and served as tests of this teaching modality. These  are used throughout the paper as reference points. 
The first event (2023) was an in-house hackathon for PhD students of the CYToskeleton as ACtive matter Research Training Group graduate school (CYTAC RTG). The second one (2024) was a public event open to all natural sciences students. Throughout the paper, we will focus on minimal goals for a simple in-house hackathon (using standard course infrastructure, e.g. for advertising) similar to the 2023 one, while incorporating relevant lessons implemented in the 2024 hackathon. A brief discussion on how to organize a more demanding outreach-oriented hackathon appears in section \ref{sec:public_hackathons}. In both of these events we taught the Cytosim software\cite{cytosim2,cytosim1} for simulating the cytoskeleton. A main motivation for teaching Cytosim was getting experimentalists and theorists, coming from multiple disciplines and backgrounds to ``speak the same language".

\partopic{Learning scope.} A crucial step in organizing a hackathon is defining goals and a scope that fit the short format. For a novice student acquiring an in-depth understanding of a simulation package can easily take a couple of months, and sometimes much more. 
Condensing this to three days is unrealistic, and a narrower scope has to be formulated; one informed also by the target audience's computational and scientific background. Our own test hackathons were designed to accommodate students with only limited programming and simulation experience (see also Fig.\ref{fig:backgrounds}). Thus, we established our scope to be: running and customizing basic simulations and setting up one simple, unique self-designed or self-modified simulation.
A secondary goal was set as: run a basic analysis of the simulation result. We prioritized this over fancier simulations, so that students can have a better grip on the full research workflow. More ambitious goals can of course be set when the participants' computational background is strong and homogeneous (for example, one of the teams in the 2024 hackathon developed a proof of concept extension to the simulation package, see blue project boxes near the end).

\begin{box_blue}
\textbf{The cytoskeleton:}
The Gö biophysics hackathons revolved around the topic of the cytoskeleton; a dynamic network of microscopic biopolymer filaments, found inside cells. It is composed of filaments such as actin and microtubules, as well as associated proteins that can for example link them or apply forces. The cytoskeleton provides cells with mechanical integrity and load-bearing capability\cite{burla2019mechanical,kechagia2023cytoskeletal}. However, unlike its macroscopic namesake, it is not a rigid structure. Instead, through a myriad of modes such as filament polymerization, severing, and molecular motor motion\cite{letort2015dynamic}, it is continuously restructuring itself\cite{burla2019mechanical}. This action-packed network also enables many morphological cellular functions, such as crawling-like locomotion and cell division\cite{bray2000cell}.
Cytosim is a highly useful tool for studying the cytoskeleton with functionality beyond standard polymer simulations, targeting the above-mentioned processes. 
\end{box_blue}
\subsection{Computational tools }
\label{sec:cytosim_cytocalc}

\begin{figure}[ht]
    \centering
    \includegraphics[width=0.15\textwidth]{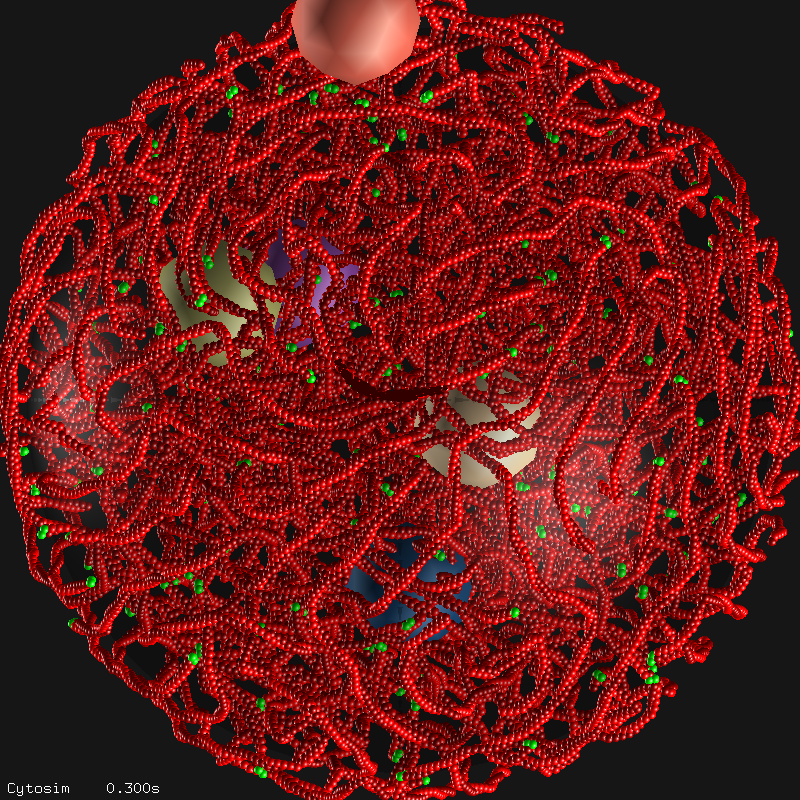}
    \includegraphics[width=0.15\textwidth]{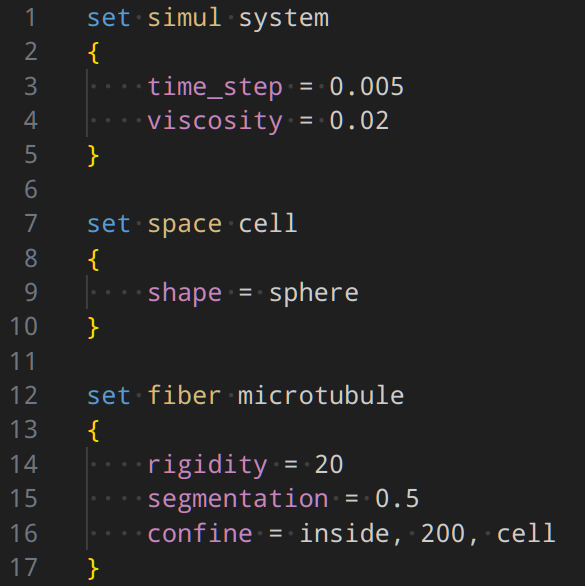}
    \includegraphics[width=0.15\textwidth]{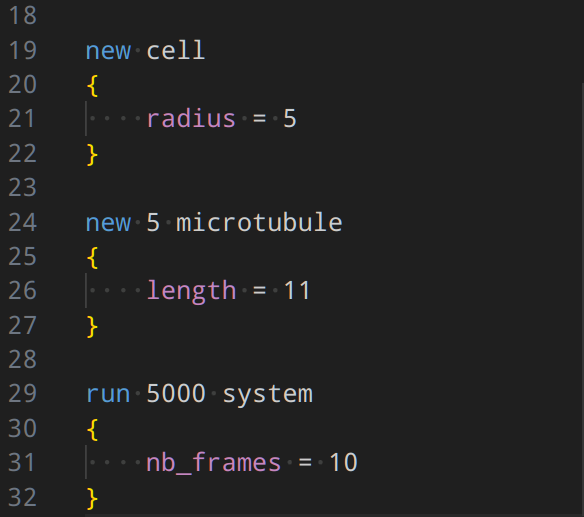}

    \caption{Cytosim simulations. Left: Example snapshot visualization. Middle and right: Minimal Cytosim configuration file taken from repository README \cite{Cytosim_READMI_repo} and used for a very first introduction to Cytosim in the hackathons. The configuration file is split for visualization. In such a script, first the various simulation objects are defined (physical properties, confinement, fibers), then objects are instantiated and finally the time progression is executed. In this example, microtubule filaments are undergoing thermal fluctuations and nothing more.}
    \label{fig:example_cym}
\end{figure}

\partopic{Cytosim.}
Not every computational tool can be learned in just a few days. Not even to the minimal functional level described above. It is thus recommended to consider various aspects such as complexity of usage, available self-learning resources, and the compatibility of these with the target audience. In our hackathons we used the Cytosim simulation package developed at the research group of Francois J. Nédélec\cite{cytosim2,cytosim1} (see Fig.\ref{fig:example_cym}). It is designed to simulate multiple flexible, and possibly elongating, filaments representing the cytoskeleton (see blue box) and various interactions between them for the sake of emulating processes in biological cells. Cytosim lowers the entry barrier significantly through detailed documentation and getting-started tutorials, and by using a simple dedicated syntax (no coding language familiarity needed). It also provides an integrated output visualization (Fig.\ref{fig:example_cym}) that helps the learner understand what their simulation is doing. 

\partopic{Cytocalc.}
To address the secondary goal in our own hackathons of analyzing simulations, we chose to supply participants with our self-developed complementary code package Cytocalc. This is because parsing of simulation output in Cytosim, such as filament positions, can present an unnecessary challenge due to its human-readable (not fully tabular) layout and lack of a standard format. Cytocalc\cite{Cytocalc_GIT_Repo} parses the output for further handling and also offers a basic set of ready-made analysis functions (such as for Mean Square Displacement and the complex shear modulus).

\subsection{The hackathon format}
\label{sec:event_format}

\partopic{Event scope.}
One thing that makes hackathons especially engaging is their intensity, with long uninterrupted daily work spans and events often ending late or lasting through the night and into the next day. Typically, to create a more immersive distraction-free experience, these are held at a secluded location, with the participants' material needs supplied on location (e.g. food, coffee and snacks). In an academic setting, financial and logistical constraints might be stricter. To lower the entry barrier also for the would-be hackathon organizers, we focus in most of this paper on in-house ``bare bones" hackathons. However, we highly recommend keeping as much as possible of the hackathon immersive atmosphere, for example by holding the event outside regular classrooms and removing ``outside world interruptions" (such as participants having to disperse for off-location lunch time). 

\partopic{Duration.} The optimal duration depends on the scientific and computational background of the target audience, and the goals set. Some points to consider are not only the actual learning of the computational tool, but also the project ideation as well as technical points such as possibly the set-up of the tool. In the 2023 Gö hackathon, teaching Cytosim to the CYTAC PhD students (who have pre-existing cytoskeleton background) worked well with two full days (although analysis was not strictly defined as a goal in this first trial).
For newcomers to the scientific topic (see also Fig.\ref{fig:backgrounds}), teaching the same tool might require longer, as was done in the three-day 2024 Gö hackathon to account for more project conceptualization time and for the added analysis goal. To simplify logistics, event days were kept discretized, and we left the option of an event ``spilling over" from one day to the next, as an exciting future experiment.

\section{Results and insights}
\label{results}

\begin{figure}
    \centering
    \includegraphics[width=0.5\textwidth]{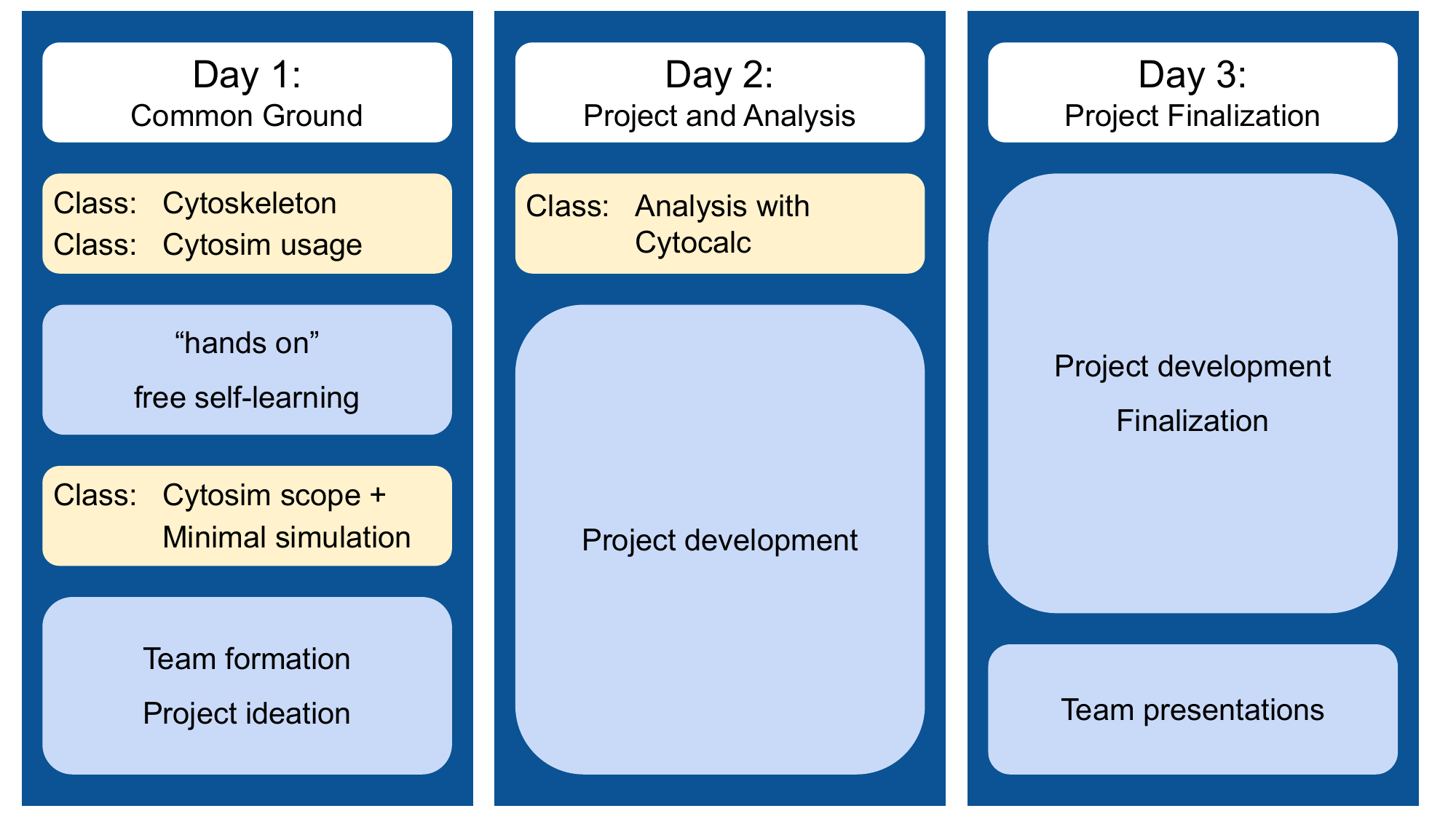}
\caption{Suggested coarse-grained schedule for a teaching hackathon (based on the 2024 Gö hackathon). The schedule is divided into `taught' sessions providing minimal usage competence (yellow), and free-form work periods for exploration and project development with only minimal individual guidance on a per-need basis (blue). The program can potentially be condensed to two days based on participant background, as explained in the text.}
    \label{fig:Programm_schematic}
\end{figure}

\partopic{Teaching scope.} 
Given that a hackathon is ultimately not a course, participants should be taught the bare minimum that allows them to start playing around on their own, even if initially with only limited understanding. A good rule of thumb is not to devote more than 10\% of the time to instructed activities\footnote{In the 2024 hackathon, only around two hours in total were devoted to core \textit{mandatory} instructed content, out of twenty-seven hours of the event. This count excludes \textit{optional} content explained in Sec.\ref{sec:public_hackathons}.}. Providing concrete instructions on how to structure the teaching in such a hackathon is unrealistic, as this highly depends on the computational tool taught. Instead, we provide a suggested coarse-grained program and guidelines for teaching Cytosim, inspired by our 2024 hackathon implementation, that the reader can potentially use as a starting point. Fig.\ref{fig:Programm_schematic} schematically shows a rough time allocation matching this program. In this example program, each day is structured to provide only the \textit{necessary} support while leaving most of the time free for independent self-charted work. It encourages participants and teams to transition from learning the basics to project conceptualization, development, and analysis, and culminates in consolidating their understanding via presentation of their results.

\partopic{Initial teaching.} The example program uses nano-classes which are modules of 20 minute teaching followed by a 10 minute Q\&A. 
The first nano-class is an overview of the scientific topic, here being the Cytoskeleton (meant for a non-expert or heterogeneous audience). This is followed by a technical nano-class demonstrating the very basic usage of the computational tool, here being Cytosim). Our own experience, and especially feedback from the 2023 hackathon, indicates that such a basic usage example is the quickest way to convey the purpose of the tool, rather than a well-thought-out and extensive introduction. 
In addition, this nano-class should also be used to motivate the participants. In our case, this was achieved in two steps. First, by demonstrating how easy Cytosim is to use, with a conceptual walk-through of using a script of an example simulation (Fig.\ref{fig:example_cym}). This was then followed up by showcasing visually the output of fancy and research-relevant simulations\cite{loughlin_computational_2010,burdyniuk_f-actin_2018} (without a detailed walk-through). After this first short teaching block, the example program already provides some ``free-play" time for participants to try things out on their own (with expert support on-hand); namely to overcome the very first minimal entry-barrier. Once students have consolidated some understanding of the basic usage, further important concepts can be introduced. Considering the minimal-teaching and maximal-self-learning guideline, this is meant more to make students aware of the relevance of these concepts, so they can explore them on their own. In our case in a second technical nano-class the students were quickly introduced to general simulation concepts such as time steps, over-damped dynamics, and thermal noise, as well as to Cytosim specific ones, such as molecular motors and cross-linkers. We concluded this nano-class again with a practical walk-through of a less trivial simulation script, pointing out where these concepts come into play \cite{Contract_tutorial}. A wider theoretical perspective of Cytosim simulations is not necessary for the intended goal, and in our case it was only provided later in the hackathon as an \textit{optional} nano-class for interested participants. 

\partopic{Team forming and project conceptualization.} 
Following this initial structured part of the academic hackathon, we suggest switching to the more traditional hackathon style characterized by unguided work. At this stage participants should have enough of an idea about the scientific topic and the computational tool taught, to start suggesting and vetting project ideas or implementation approaches, and can be encouraged to form teams. 
In our hackathons we gave complete freedom in team forming, and opted to leave the project ideation process fully to the participants, rather than provide pre-defined projects. However, organizers were on-hand to provide support regarding relevance, importance and scope of envisioned projects as well as for technical questions. 

Although the hackathon methodology is aimed at maximal learning freedom, teams can still benefit from the more experienced perspective of the organizers. While organizers should be careful not to accidentally lead teams in a given direction, they could for example provide technical feedback on the feasibility of a given implementation approach. Indeed, in our experience it is good practice for organizers to routinely check up on teams throughout the hackathon. Due to the short time format, teams might have to pivot their projects even for rather simple implementation challenges, that otherwise could be overcome.

\partopic{Beyond the entry barrier.} 
In principle, the rest of the time could be dedicated to self-learning via the teams' projects. However, as stated above, in our own hackathons a secondary goal was to get teams to analyze the results of their simulations. For this reason, our example program includes a last mandatory nano-class on analysis using Cytocalc (see Sec.\ref{sec:cytosim_cytocalc}). This was offered only on the second day to allow participants to first consolidate some understanding of Cytosim. Offering it early in the morning, was aimed at encouraging teams to think ahead, and adapt the simulations to a desired end-product while still under development and easily changeable. Finally, following the longest part of the hackathon devoted to independent project development, we suggest that teams should present their projects briefly. This is meant to consolidate their understanding, to learn from each other's experience, and to provide the organizers with a good overview of the learning done. Two example projects presented in the 2024 hackathon are provided in the blue boxes towards the end.

\begin{table*}
\caption{Comparison of 1st and 2nd trial hackathons.}
\begin{tabular}{@{}lll}
\hline
 & 1st event (2023) & 2nd event (2024) \\ 
\hline
Participants & 10 & 18 \\ 
Biophysics familiarity & experts & heterogeneous \\
Programming experience & limited, outside of courses* & wide range \\
Academic level & PhD students & Bachelor to PhD Students \\
Duration & 2 days & 3 days\\
Daily timeline & 09:00-16:00 & 09:00-19:00\\
Goals & Cytosim simulations & Cytosim simulations + analysis\\
Advertising & in-house & multi-channel\\
Venue & classroom + computer lab & ``secluded" venue @ MPI-NAT\\
Computational hardware & Personal laptops + computer lab & Centrally supplied laptops
\\
\hline
\end{tabular} \\
* With two exceptions.
\label{table:HackCompare}
\end{table*}

\subsection{From amazingly minimal to minimally amazing: in-house vs. public hackathons}
\label{sec:public_hackathons}

 So far we've focused on in-house hackathons meant to teach a computational tool useful for the target audience. In a public hackathon however, the benefits are oriented more towards outreach, promoting multidisciplinary knowledge transfer, recruiting students and networking, and the computational tool taught is used as a means to those ends.

 In comparison to in-house events where organizational efforts can be reduced to a minimum, public events are more demanding, not the least for the sake of drawing applicants. Further effort has to be expended on providing minimal scientific background that external students might be lacking. Finally, to match participant expectations from such a public event, it is advisable to adhere more closely to the original hackathon spirit and ensure better the immersive atmosphere, such as via a secluded location, an encapsulated environment, and possibly a prize for the winners. 

 This approach was tested in the 2024 hackathon, where we aimed to help students cross entry barriers to biophysics research. Advertisements for the event required significant efforts and planning ahead, with admittedly much room for improvement (e.g. due to suboptimal timing in the semester break). Scientific background was supplied (beyond the aforementioned first mandatory nano-class overview of the cytoskeleton topic) via further \textit{optional} nano-classes on the cytoskeleton, offering a variety of discipline perspectives. The optional nano-classes also doubled as a way to attract applicants, by offering classes on buzzy topics such as AI-coding. One should note that we do not count these extra nano-classes towards core teaching time, and explicitly advised participants \textit{not} to attend all of them.
To make the event more immersive, the 2024 hackathon was held away from the usual classrooms (at one of the event rooms of the nearby MPI-NAT: Max Planck Institute for Multidisciplinary Sciences), and worldly needs such as lunch, coffee, beverages and snacks were provided to participants on location. Finally, a prize was offered to winning teams, sponsored by the Alumni Göttingen e.V.

\subsection{Evaluation, insights and products}
We now turn to evaluating the results of the two Gö biophysics hackathons, based on both qualitative participant feedback (written and oral) as well as polls conducted during the 2024 hackathon. We examine the participant backgrounds, insights on the self-learning and participant-reported learning outcomes. Finally, we provide a glimpse into the teams' projects. 
\begin{figure}[ht]
    \centering
    \includegraphics[width=\linewidth]{"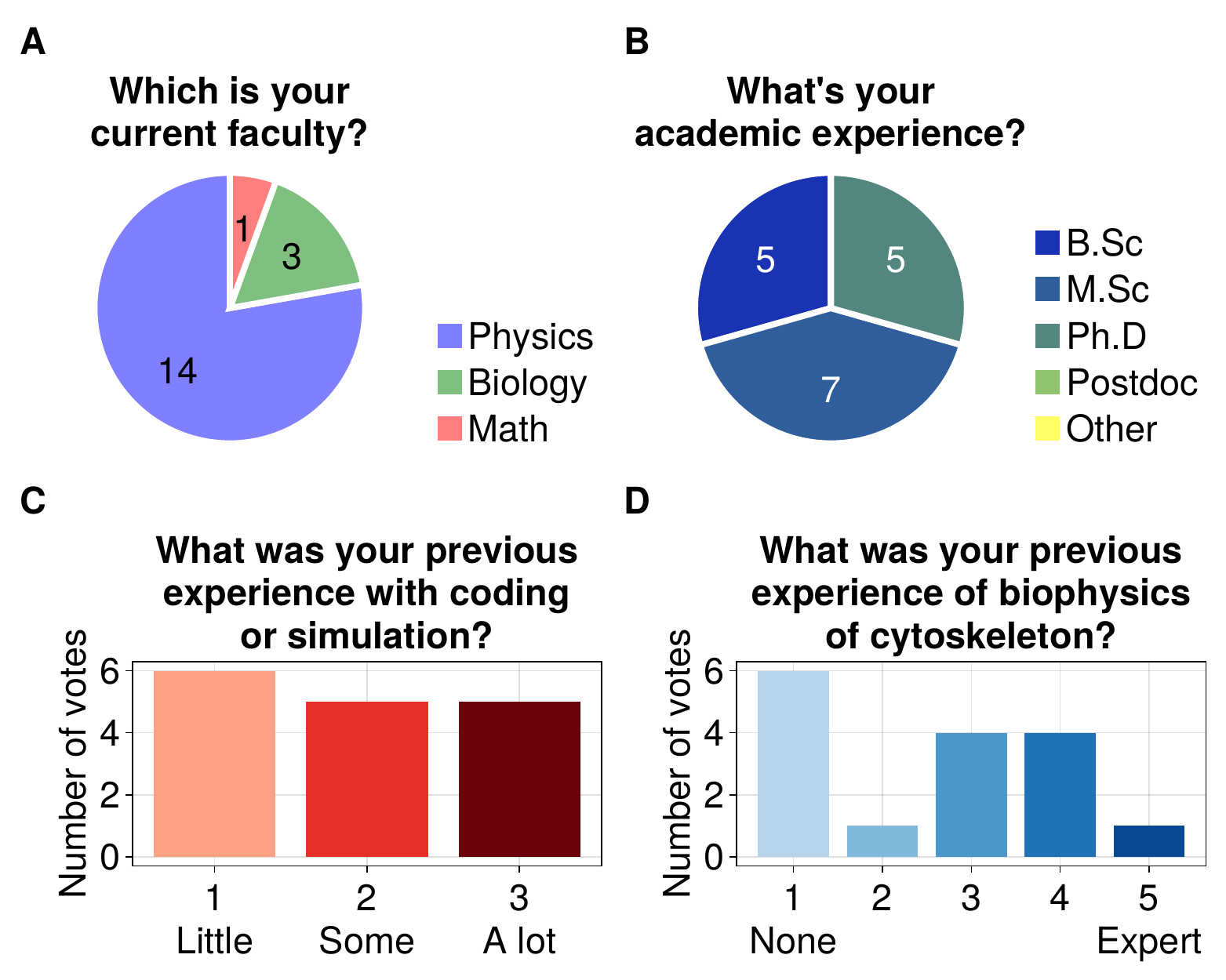"}
    \caption{Diversity of participant scientific and technical background in the 2024 hackathon (respondent number given in curly brackets). The first two panels were obtained in an informal Whatsapp poll, while the last two come from an anonymized exit poll.
    \textbf{A} The academic programs' association \{18\}.
    \textbf{B} Academic experience \{17\}.
    \textbf{C} Previous experience with coding and simulations \{11\}.
    \textbf{D} Previous experience with biophysics and cytoskeleton \{16\}.}
    \label{fig:backgrounds}
\end{figure}

\partopic{Background of participants.}
\label{sec:results}
The 2023 and 2024 hackathons were aimed at different target audiences, as mentioned above (and see also Tab.\ref{table:HackCompare}). In 2023 participants were the CYTAC RTG PhD students working on cytoskeleton in the fields of experimental biophysics, chemistry, cell-biology, statistical physics, and mathematics. Around 70\% of the participants had only limited coding experience outside academic courses (with two notable exceptions), but they all had already a very good knowledge of the cytoskeleton. In both these respects, the participants' background in the 2024 \textit{public} hackathon was more heterogeneous (see Fig.\ref{fig:backgrounds} panels C and D). Also in terms of the academic stage, students spread evenly from B.Sc to Ph.D, as shown in panel B. An unintended bias in the advertisement efforts lead in turn to a bias towards physics in the participants' scientific discipline composition for that event (panel A).

\partopic{Qualitative learning insights.}
With regards to participants' learning approaches, we observed  three different modes:  1. instructed learning (in the nano-classes or individually upon request), 2. following  tutorials and examples offered on the Cytosim repository, and 3. free-from trying out (and then learning why things fail either on their own, from teammates, or by asking instructors). For the instructed content, as mentioned before participant feedback from the 2023 hackathon suggests that  the workflow for using simulations can be grasped quicker when starting with practical, hands-on  usage instruction before introducing theory and general concepts of simulations. This approach  was implemented in the 2024 hackathon, and based on participant interactions  (e.g. in-class and follow-up questions ) indeed streamlined the initial on-boarding. The \textit{optional} nano-classes offered in 2024 (see section\ref{sec:public_hackathons}) provided some complimentary theory, with the explicitly stated aim of transferring knowledge not uniformly to all, but rather more organically from instructor to interested team member and then potentially diffusing to the rest of the team. However, we do not consider this methodology as a core part of the hackathon format, and its usefulness remains to be evaluated.

 Following written tutorials helped participants grow their knowhow beyond the level of roughly understanding the single script already taught, while allowing hands-on practice and some freedom of experimentation. These tutorials and example configuration files based on published scientific work also exposed the participants to current cytoskeleton research and how Cytosim can be used to solve real-world research questions. Such tutorials can also provide a starting point for project ideation, and indeed some of the teams came up with projects that constituted follow-ups to tutorial problems.

 In terms of time spent, by far the largest portion was free-form learning, engaged in already during the first day, and for almost all of the next days. Although it is our impression that the high motivation of the teams for their projects made this mode highly effective, the extent of the learning done in this way is harder to gauge. This is because each team and participant learned different aspects of Cytosim and Cytocalc. Another reason is that a lot of the learning was done by making and then correcting mistakes and by running up into implementation dead ends, which were reported only sporadically, and almost never made it to the final presentations. While all of the three mentioned learning modes are of course also used in standard courses, a unique feature of this hackathon is that participants can switch between them repeatedly and without delay.  

\partopic{``Blue-sky" format.} As already mentioned, in both of our hackathons we opted for an open-ended style (``Blue-sky"), without supplying teams with pre-defined projects. The reason for this choice is that we personally see great value in the project-ownership that comes from teams developing their own project concept. While this was a natural choice in the 2023 hackathon due to participants' scientific background, it was less obvious in the public hackathon. Indeed, the Blue-sky format was initially met with lack of self-confidence in participants' ability to come up with a project that makes sense. However, with encouragement and occasional feedback from the organizers, this hurdle was overcome and in both hackathons all teams managed to conceptualize a project. Considering the final project presentations, we consider this a success, and note also positively the variety of project themes  as a learning opportunity for the organizers. 

\begin{figure}[ht]
\centering
    \includegraphics[width=\linewidth]{"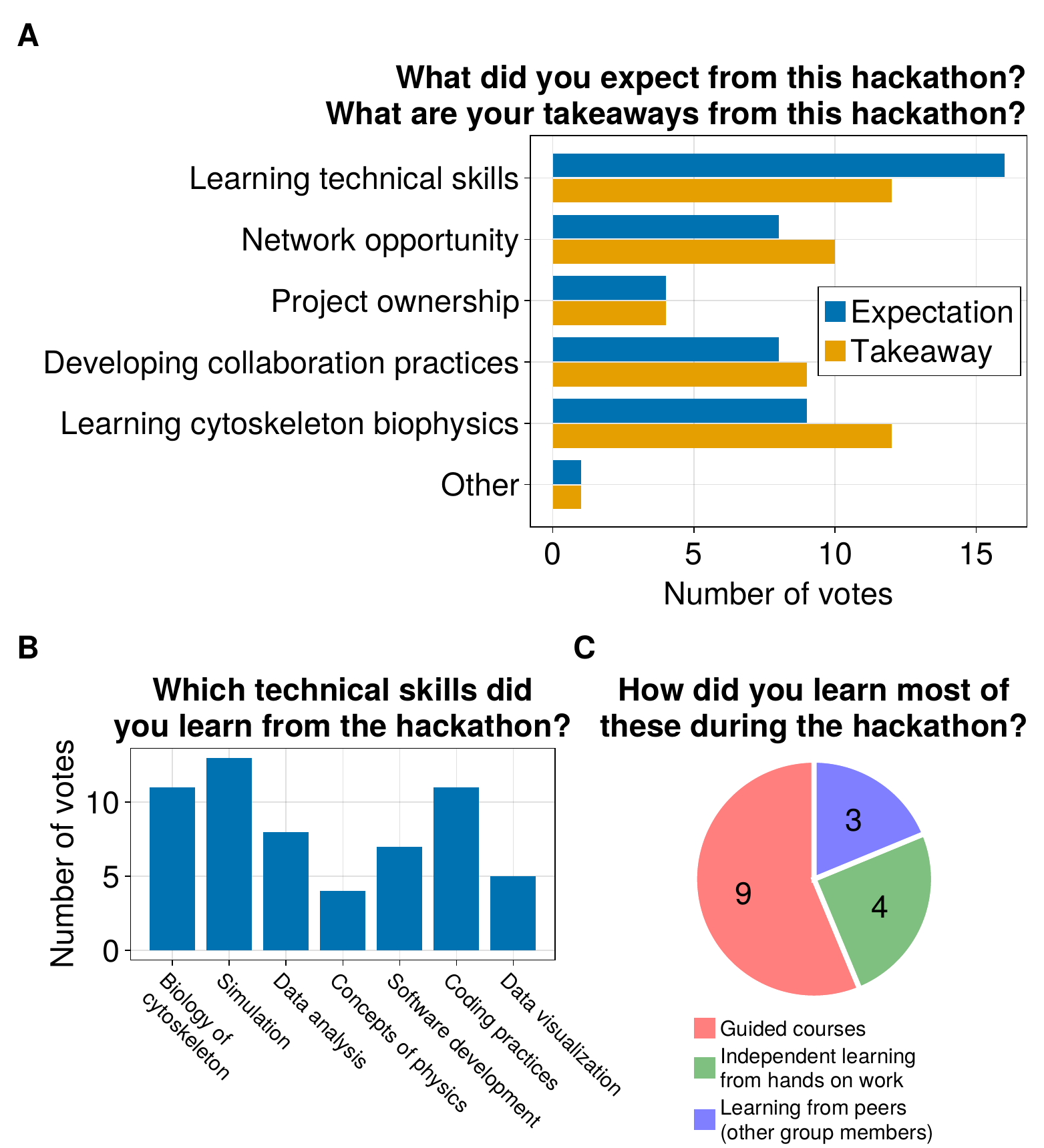"}
    \caption{Learning outcomes and expectations based on a participant self-assessment exit poll in the 2024 hackathon (16 respondents). Results in \textbf{A} and \textbf{B} correspond to multiple choice questions. 
    \textbf{A} Comparison of the participants' primary expectations (blue) and takeaways (yellow) from the hackathon. 
    \textbf{B} Technical skills learned at the hackathon. 
    \textbf{C} Participant perception of how most of these technical skills were learned.}
\label{fig:Expectation}
\end{figure}

\partopic{Quantitative evaluation.} To gain also a somewhat more objective measure of the participants' learning experience, we employed an anonymized exit poll, whose results are shown in Fig.\ref{fig:backgrounds} (bottom) and Fig.\ref{fig:Expectation}. Three questions were in the multiple choice answer format (Fig.\ref{fig:Expectation} panels A and B), while the others were selection list questions. We asked students first which benefits they expected to obtain from the hackathon (see possible answers in Fig.\ref{fig:Expectation} panel A). Students were then asked which of these benefits were indeed obtained (same panel, takeaway columns). 
We see that beyond the expected gain of technical knowledge, most participants also report to have benefited in other respects, such as networking and learning about the cytoskeleton.

We further asked about specific technical skills learned (see possible answers in panel B). In this regard, we see that simulation concepts were the most dominant. This, together with the impressive team projects (see blue boxes), and most of the participants' statement on lack of significant simulation experience (Fig.\ref{fig:backgrounds}, panel C),  reassures us that indeed the hackathon format can be effective in teaching simulations and specifically Cytosim. Besides  the various  computational skills learned or developed further, such as Coding practices and Data analysis, we note the high number of respondents that reported learning about Biology of the Cytoskeleton. We consider this  evidence for the advancement of multidisciplinary, considering the Physics and math background of most participants (Fig.\ref{fig:backgrounds}
panel A).

Finally, we asked students about modes of learning in the hackathon, and specifically about how they learned the most of the skills they reported on in panel B. Here we see a definite majority for the guided courses (nano-classes). Nevertheless, the main insight that one could extract from this result is that multiple modes of learning were used during the hackathon (see argument below).

\textbf{Evaluation limitations.}
In addition to inherent limitations of poll-based evaluation, we note a few possible improvements. The strong correlation between expectation and takeaway (Fig.\ref{fig:Expectation} is likely influenced by both questions appearing in the exit poll. A better approach would be to poll expectations beforehand. The question assessing learning in Fig.\ref{fig:Expectation} C was in retrospect too broad to specifically evaluate the core Cytosim learning, as it refers to \textit{all} technical skills mentioned in the previous question. Furthermore, the 2024 hackathon included many off-topic \textit{optional} nano-classes, which were grouped under ``guided courses", making it difficult to distinguish their impact from (the few) core mandatory classes. Also, participants, being accustomed to guided courses, may have attributed these a greater importance. Future improvements include incorporating self-assessment on specific Cytosim concepts, employing more objective learning assessments (not poll-based) while balancing low-stakes principles that exclude official grading, as well as   using control groups to refine our evaluation process.

\partopic{Projects.} For assessing the effectiveness of the hackathon format, it is useful also to briefly review the resultant team projects. In total, eleven groups were formed in the two hackathons, with projects mostly focused on biophysics-oriented problems as well as a couple of projects focused on Cytosim software solutions. A couple of project examples are given in the blue boxes. Projects in the 2023 hackathon were largely inspired by the experiments run by the PhD students attendees, while projects in the 2024 hackathon either expanded on themes introduced in tutorials on the Cytosim website, or were completely innovative. Projects typically used Cytosim scripts that were not a mere copy of an existing example with only changed parameters, but rather included additional/altered functionality. These projects thus demonstrate concrete understanding of the Cytosim workflow. Analysis of simulation results was performed in most projects, to varying degrees of complexity, demonstrating learning outcomes starting from understanding of the general simulation research workflow to technical mastery of the provided Cytocalc analysis package.

\begin{box_blue}
{\centering Project example 1 \par}
\textbf{im2cym:}
A software implementation for converting microscopy images into an initial state of a Cytosim simulation. A set-up simulation stage is added where filaments ``stick" in place based on the supplied design.
This image-matching state may then be used as the starting point for simulations to study realistic cellular structures; addressing a hitherto unrealized functionality gap.
\begin{figure}[H]
    \centering
    \begin{subfigure}{.31\textwidth}
        \includegraphics[width=\textwidth]{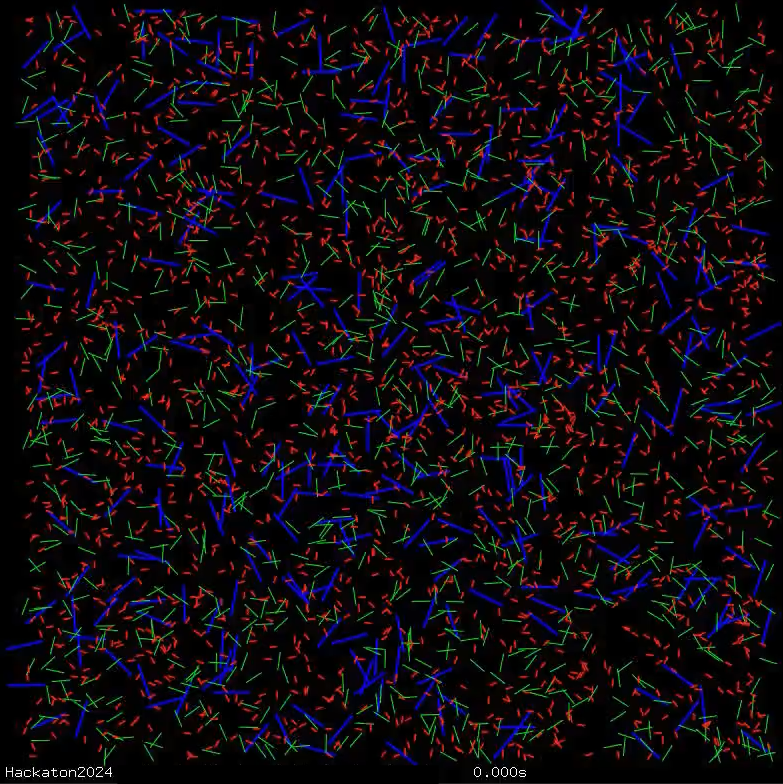}
    \end{subfigure}
    \begin{subfigure}{.31\textwidth}
        \includegraphics[width=\textwidth]{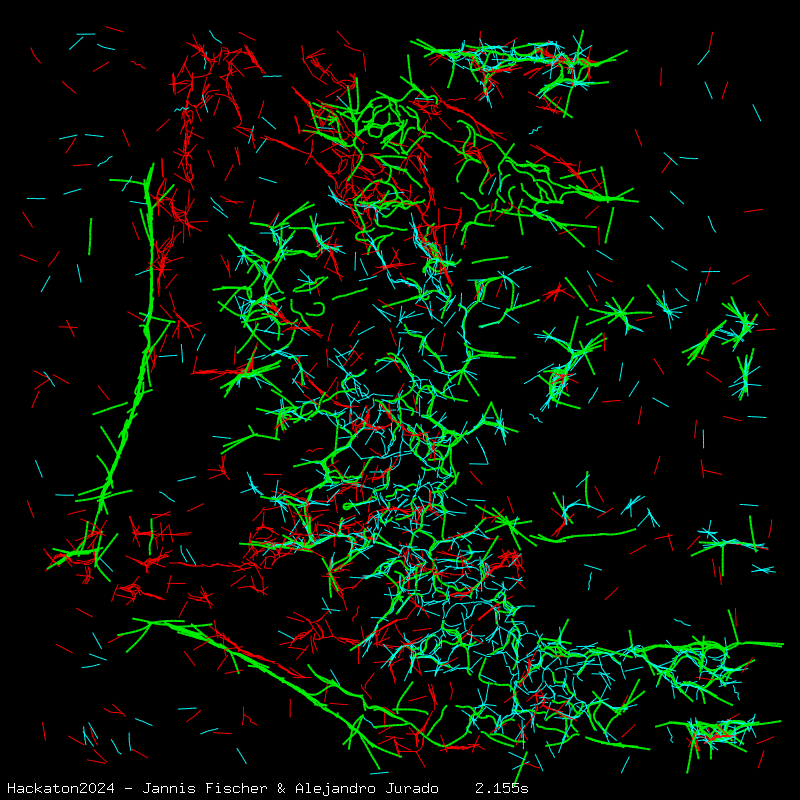}
    \end{subfigure}
    \begin{subfigure}{.31\textwidth}
        \includegraphics[width=\textwidth]{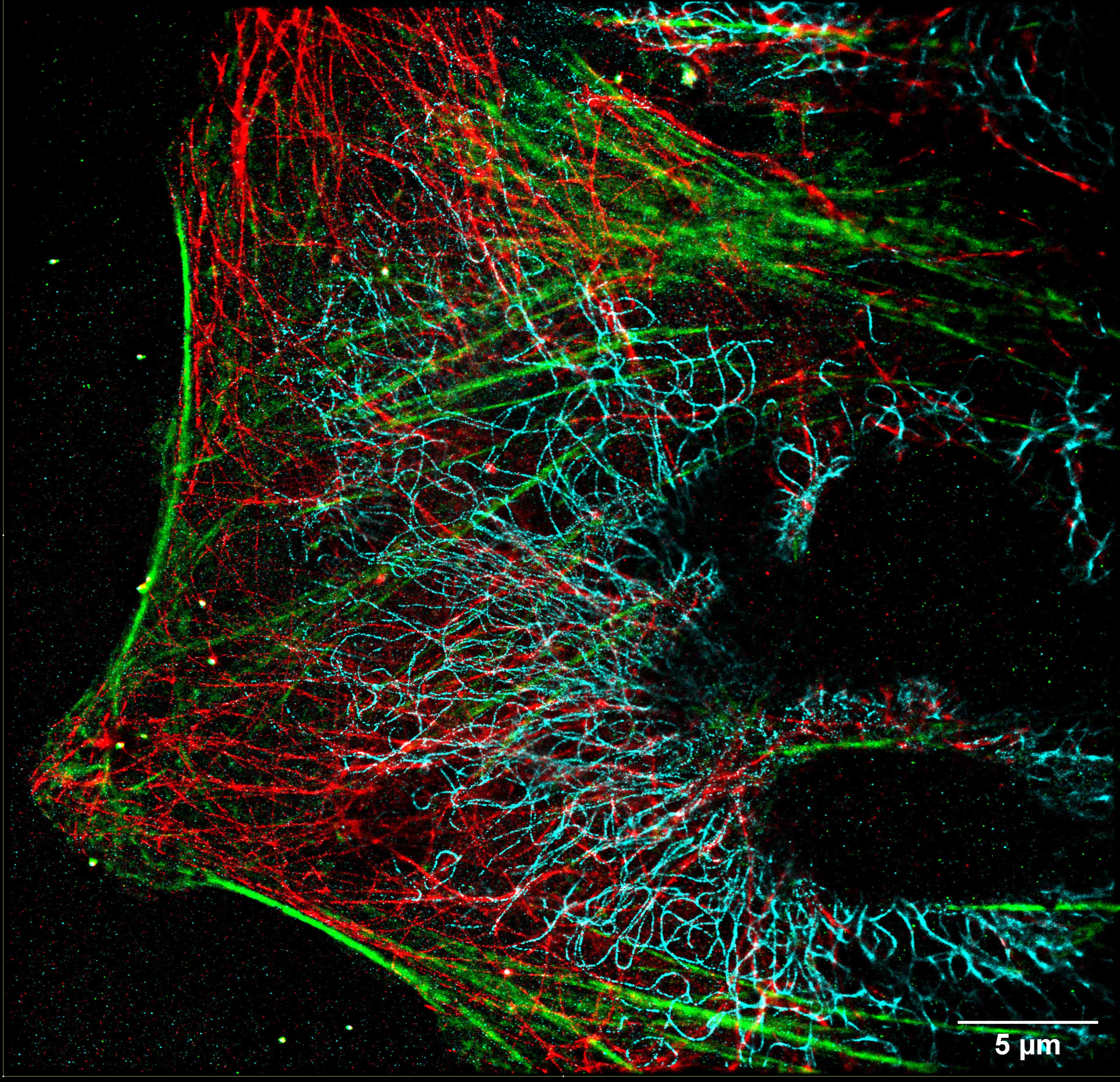}
    \end{subfigure}
    \caption{Left and middle: the \texttt{im2cym} added set-up stage of the simulation (initial and final states respectively). Right: The multiplexed super-resolution image provided as input, courtesy of Dr. Roman Tsukanov, Göttingen Third Institute of Physics.}
    \label{fig:im2cym-result}
\end{figure}
\end{box_blue}

\section{Discussion}
 In this work we proposed the concept of a teaching-hackathon as a method for overcoming computational entry barriers and we tested it for teaching the simulation software Cytosim, used for cytoskeleton research. In two hackathons organized by us in Göttingen campus in 2023 and 2024, we showed that basic usage of such a simulation software could be learned in 2-3 days, and that learning can happen with the envisioned minimal guided teaching followed by self-driven independent learning, once participants passed the entry barrier. We incrementally improved the suggested format between the two trials by reducing the amount of structured teaching. Based on organizer and participant feedback, we focused the teaching on practical knowhow, while keeping the theoretic material (even basic concepts) as \textit{optional} in-hackathon teaching or for a potential later followup learning.

\begin{box_blue}
{\centering Project example 2\par}
\textbf{Anaphase chromosome separation:}
A minimal model simulating anaphase during cell division. 
Microtubuli are capturing and separating chromosomes. Parameters related to filament growth and interaction with chromosomes were studied for their effect on efficiency and correctness of chromosome separation.
\begin{figure}[H]
    \centering
\includegraphics[width=.31\textwidth]{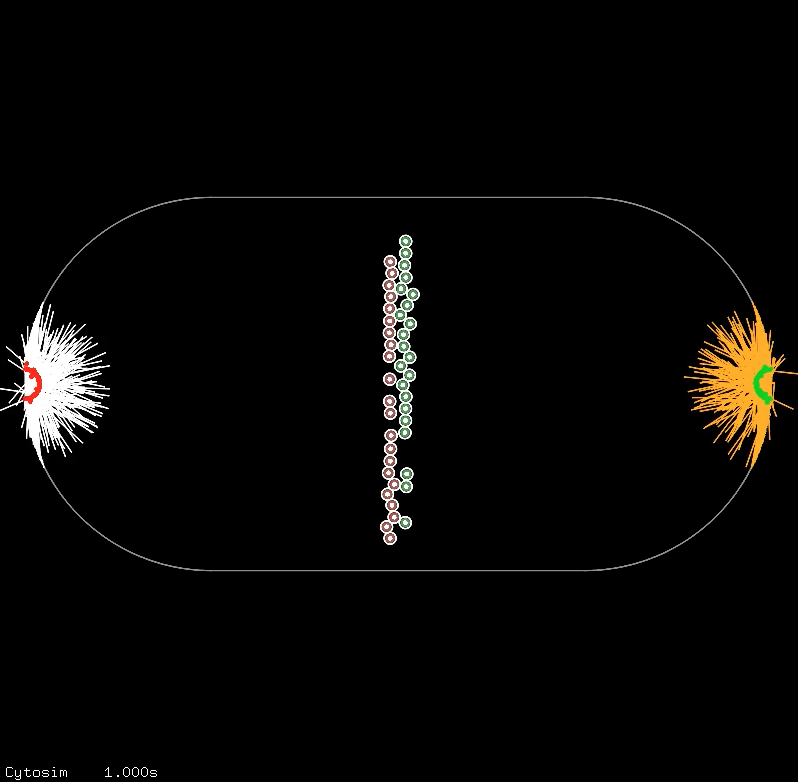} 
\includegraphics[width=.31\textwidth]{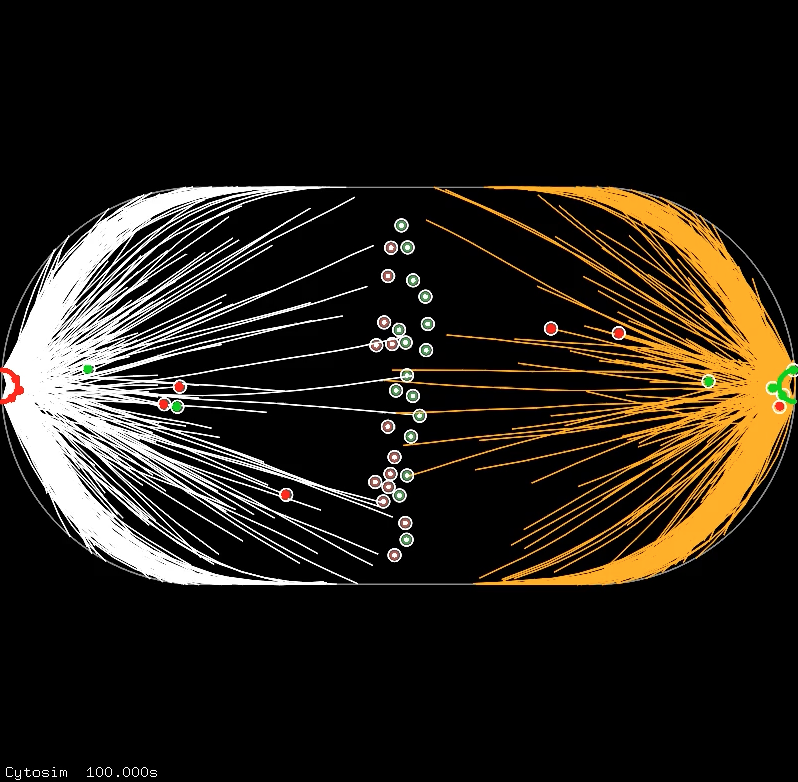}
\includegraphics[width=.21\textwidth]{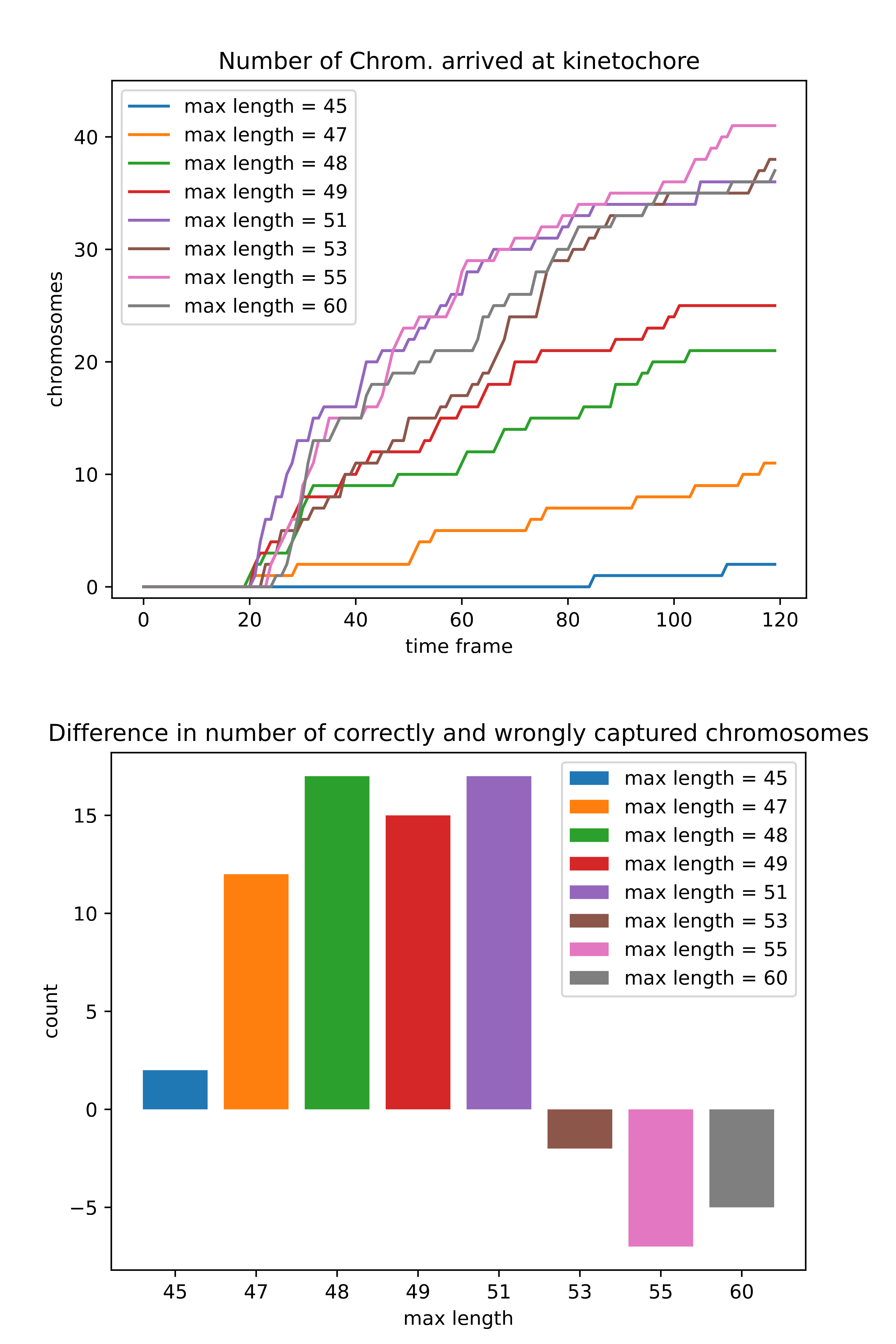}
    \caption{Left and middle: Initial and interim states of simulated chromosome capture in cell anaphase. Right: example analysis of optimal microtubule parameters for cell division.}
    \label{fig:mitosis-result}
\end{figure}
\end{box_blue}

Such a teaching methodology aiming at entry barriers is especially useful in interdisciplinary fields such as biophysics, where students often come from diverse backgrounds both in terms of scientific disciplines and in terms of computational experience. Specifically, while programming courses are practically standard in natural science degrees, it is a big leap from completing course exercises (where a known solution is only a Google-search or an AI-prompt away) to confidently exploring open problems. Indeed, we showed that the hackathon format works both for a target audience heterogeneous in its scientific training (2023 hackathon) and its computational experience (2024).
 
 Furthermore, this work shows the versatility of use cases for the hackathon format. Whereas most non-academic hackathons focus on improving existing computational skills\cite{briscoe2014digital}, we show here that the format can be used for teaching new computational tools and skills, countering assumptions to the contrary proposed in previous studies\cite{schulten2024we,gama_hackathon_2018}. The format is also versatile in its potential end-products, which can either follow pre-defined specifications or be fully open-ended, although in this work only the ``Blue-sky" format was explored.
 
We observe that an unexpected key challenge we faced as organizers was overcoming our  teaching tendencies ingrained in us during our own academic training.  Indeed, despite aiming for minimal teaching, feedback in 2023 showed that this objective was not met. In 2024 however, we trusted the hackathon format much  more to encourage independent learning rather than relying on structured teaching. Another challenge revealed through participants' feedback was  to adjust our perspective from a presumably theorist's approach to one that also accommodated experimental-focused students, and promote learning by first providing concrete examples and only later strengthening it with theory and general principles.
 
\partopic{Evaluation.} We further conducted preliminary evaluation of the hackathon format, regarding benefits beyond the minimal core learning outcome. Our evaluation shows students gained many additional benefits, both in terms of developing supplementary computational skills, developing collaboration skills, networking and learning about topics outside their discipline.

In terms of evaluating the success and effectiveness of learning of the core topic, rigorous evaluation is still lacking, and in fact the absence of accepted evaluation criteria for the format is apparent in previous demonstrations of the format\cite{oyetade2022educational,mhlongo2020effectiveness, pakpour2022engaging} and was already pointed out to in a previous comparative study\cite{schulten2024we}. The presented team projects were suggested in some of these studies as a possible evaluation criteria, but we point out that this shifts the focus away from the learning goal. Indeed, in two cases in our hackathons team end-products were deemed ``insufficient" (by the team itself or by the organizers) and yet  involved significant learning.
Perhaps most importantly, we point out that the limitations of the format have not been rigorously explored here: which computational tools are not adequate to be taught ``from scratch" in so little time. Indeed, Cytosim represented an almost optimal choice, combining a user-friendly interface (no extensive programming experience needed), rich documentation and  and built-in visualization. Furthermore the scalability of the format should be tested: whether hackathons can be integrated into academic training programs, and how to retain their appeal in that scenario.

\partopic{Teacher benefits.} While teaching is typically oriented to benefit the learners, the hackathon format can carry with it also benefits for the teachers and organizers, which should encourage the reader to try it themselves. These include introducing into the research innovation from external sources (e.g. the Im2Cym project), student recruitment, networking also outside of one's discipline and outreach.

Finally, it is worth situating the hackathon format within the broader landscape of academic teaching. Rather than replacing traditional teaching methods, hackathons offer a complementary and innovative approach that supports independent problem-solving and interdisciplinary collaboration. They can serve as an effective bridge between structured coursework and real-world scientific challenges, preparing students to engage with research in a self-directed and collaborative manner.

 \section{Author Contributions}
We note that the number of authors by no means reflects the necessary number of hackathon organizers, only the authors' collaborative tendency, and that especially in-house hackathons can be organized even by one or two people.

YGP: Designed teaching concept, designed teaching materials, curriculum, and evaluation methods, performed teaching, and wrote the manuscript with contributions, review and feedback from KB, PZ, RM and EK.
KB: Designed teaching materials and curriculum, performed teaching and provided participant support, analyzed evaluation data, literature research, created data visualizations, and contributed to writing the manuscript.
AH: Designed teaching materials, performed teaching and provided participant support, data collection.
EK: Performed teaching, managed communications with participants, and contributed to writing the manuscript.
RM: Performed teaching, contributed to curriculum design and contributed to writing the manuscript.
KIVS: Designed teaching materials, performed teaching and provided participant support.
GW: Performed teaching, provided participant support and literature research.
PZ: Performed teaching and provided participant support, created data visualizations and contributed to writing the manuscript.

\section{Acknowledgements}
    Financial and logistic support for the hackathons was provided by the  the Deutsche Forschungsgemeinschaft (DFG, German Research Foundation) – Project-ID 449750155 – RTG 2756, projects A3, A4, A5 and B4, and by Alumni Göttingen e.V.
    
    The authors thank all additional nano-classes instructors, who very kindly agreed to enrich the hackathon experience with their expertise:  H. Bruns, M. Eskandari, Prof. Dr. S. Klumpp, M. Thomas, Dr. R. Tsukanov, T. Weege.   Special thanks also to J.Singh and A. Smagliuk, for both teaching and providing a decisive contribution to the hackathon organization.    
    Finally, the authors would like  to thank Dr. P. Lenart for his crucial support and to Dr. P. Klein for his critical manuscript feedback.

\printbibliography
\end{document}